\documentclass[12pt]{article}
\DeclareUnicodeCharacter{0301}{\'{e}}
\usepackage{amssymb}
\usepackage{amsmath}
\usepackage{amsthm}
\usepackage{color}
\usepackage{comment}
\usepackage{enumitem}		% for fancy lists
\usepackage{geometry}		% allows easy margin settings
\usepackage{graphicx}
\usepackage{authblk}
\usepackage{amssymb}
\usepackage{listings}
\usepackage{xcolor}
\usepackage{caption}
\usepackage{color}
\usepackage{algorithm}
\usepackage{algpseudocode}
\usepackage{booktabs}
\usepackage{subcaption}

\makeatletter
	\@addtoreset{equation}{section}
\makeatother

\let\originalleft\left
\let\originalright\right
\renewcommand{\left}{\mathopen{}\mathclose\bgroup\originalleft}
\renewcommand{\right}{\aftergroup\egroup\originalright}

\newlist{romanlist}{enumerate}{3}
\setlist[romanlist]{label=\roman*),ref=(\roman*)}

\begin{document}

\newcommand{\cF}{\mathcal{F}}
\newcommand{\cP}{\mathcal{P}}
\newcommand{\cR}{\mathcal{R}}
\newcommand{\cS}{\mathcal{S}}
\newcommand{\cT}{\mathcal{T}}
\newcommand{\ee}{\varepsilon}
\newcommand{\rD}{{\rm D}}
\newcommand{\re}{{\rm e}}

\newtheorem{theorem}{Theorem}[section]
\newtheorem{corollary}[theorem]{Corollary}
\newtheorem{lemma}[theorem]{Lemma}
\newtheorem{proposition}[theorem]{Proposition}

\theoremstyle{definition}
\newtheorem{definition}{Definition}[section]

% MSC codes:
%		37G35 -- Attractors and their bifurcations
%		39A28 -- Bifurcation theory

\title{
Deep Learning and Chaos: A combined Approach To Image Encryption and Decryption
}
\author[1]{Bharath V Nair}
\author[1]{Vismaya V S}
\author[1]{Sishu Shankar Muni}
\author[2]{Ali Durdu}
\affil[1]{School of Digital Sciences,\\ Digital University Kerala\\
Thiruvananthapuram, PIN 695317, Kerala, India}
\affil[2]{Department of Management Information Systems, Faculty of Political Sciences, Social Sciences University of Ankara, Ankara, Turkey}

\maketitle

% keywords:
% MSC codes:

\begin{abstract}
In this paper, we introduce a novel image encryption and decryption algorithm using hyperchaotic signals from the novel 3D hyperchaotic map, 2D memristor map, Convolutional Neural Network (CNN), and key sensitivity analysis to achieve robust security and high efficiency. The encryption starts with the scrambling of gray images by using a 3D hyperchaotic map to yield complex sequences under disruption of pixel values; the robustness of this original encryption is further reinforced by employing a CNN to learn the intricate patterns and add the safety layer. The robustness of the encryption algorithm is shown by key sensitivity analysis, i.e., the average sensitivity of the algorithm to key elements. The other factors and systems of unauthorized decryption, even with slight variations in the keys, can alter the decryption procedure, resulting in the ineffective recreation of the decrypted image. Statistical analysis includes entropy analysis, correlation analysis, histogram analysis, and other security analyses like anomaly detection, all of which confirm the high security and effectiveness of the proposed encryption method. Testing of the algorithm under various noisy conditions is carried out to test robustness against Gaussian noise. Metrics for differential analysis, such as the NPCR (Number of Pixel Change Rate)and UACI (Unified Average Change Intensity), are also used to determine the strength of encryption. At the same time, the empirical validation was performed on several test images, which showed that the proposed encryption techniques have practical applicability and are robust to noise. In contrast to most existing encryption algorithms, which focus on single-image encryption, our method efficiently combines techniques from hyperchaotic maps and deep learning to improve encryption efficiency and robustness. Simulation results and comparative analyses illustrate that our encryption scheme possesses excellent visual security, decryption quality, and computational efficiency, and thus, it is efficient for secure image transmission and storage in big data applications.
\end{abstract}

\section{Introduction}
Chaos contains highly unpredictable characteristics used in many fields \cite{guan2005}. In securing digital information against unauthorized access and tampering, the inherent unpredictability and complexity of the chaotic systems provide a robust framework \cite{luo2014}. On image encryption and decryption, chaos can efficiently generate intricate sequences to scramble pixel values. This will ensure that an encrypted image is just noise to the naked eye, and, computationally, it would be infeasible to reconstruct it without the decryption key \cite{shafique2021}. This nonlinear dynamics feature is used by chaotic maps, such as 3D hyperchaotic maps and 2D memristor maps \cite{bao2021}, to provide high security for the encrypted data through their large dimensionality and complex trajectories. Cyber-attacks and confidential data loss are very common nowadays, so hiding sensitive information from unauthorized parties is more necessary. Hence, encryption techniques are important. The more recent techniques available in this regard for image encryption are mostly cryptographic algorithms, which utilize mathematical transformations, substitution-permutation networks, or chaotic maps \cite{li2019}. This ensures the confidentiality and integrity of the images since plaintext images are converted into ciphertext, which can only be decrypted using some specific key. More on this, audio encryption \cite{babu2021} can be done by applying chaotic sequences or some cryptographic algorithms to the audio data that is to be transmitted or stored, hence protecting privacy from access by unauthorized people \cite{hamouda2020}. Deep learning-based approaches \cite{ding2021} changed the notion of traditional encryption techniques in the very recent past with the introduction of Convolutional Neural Networks (CNNs) \cite{alzubaidi2021} in this field. They can learn complex patterns and features in images, which can be applied to key management, anomaly detection, and optimizing encryption algorithms. Only this functionality improves the efficiency of the processes of encryption and resiliency in case of possible attacks \cite{lin2024}. Pseudo-random number generators \cite{hu2020} play an important role in generating unpredictable sequences applied for encryption keys and chaotic maps. PRNGs ensure secure communication or cryptographic operations by creating sequences that seem random but would be generated deterministically from a seed of value. Two of the most demanded features from any securely done encrypted data are retaining its confidentiality and integrity, quality, and non-predictability of the PRNG outputs \cite{gutub2021}. Effective encryption mechanisms guarantee secure communication so that no interception of information takes place in any form. Therefore, Secure communication protocols leverage robust symmetric and asymmetric cryptography techniques, key management schemes, and authentication mechanisms to ensure that only the communicating ends can effectively read and understand sensitive information exchanged over a network. Deep learning models can be used to strengthen these algorithms while optimizing their performance in real-world applications. Deep learning models have dedicated automated feature extraction, pattern recognition, and neural network training, hence developing much faster schemes of encryption that secure large-scale data. The main purpose of the proposed algorithm is to provide security to important images from unauthorized people and prevent them from being misused. It is done through image encryption.

This paper presents a detailed study on implementing and testing an image encryption and decryption technique using a 3D hyperchaotic map and a 2D memristor map integrated with CNN. We first looked at the theoretical background of chaotic systems, their behaviour, and the importance of Lyapunov exponents \cite{ding2023} in 3D hyperchaotic maps. We found that the complex dynamics of 2D memristor maps were used to scramble image pixels. To ensure high security, the problem of image encryption was solved by using the complex sequences generated by these chaotic maps. We took images, converted them to NumPy arrays, and performed the encryption process using bitwise XOR \cite{kumar2011} and modular arithmetic. In addition to using chaotic maps, CNN enhanced the decryption process by being trained to effectively recreate the original images from their encrypted counterparts. We checked for the robustness of our algorithms through security analyses; entropy analysis\cite{farhan2019} for a measure of randomness, correlation analysis \cite{singh2019} to check the linear relationship between the neighbouring pixels, and anomaly detection to check the difference between the original and decrypted images. Further, histogram analysis was done to study the distribution of pixel intensities \cite{mohammad2017}, and robustness to noise analysis was done to study the algorithm's resistance to Gaussian noise \cite{liu2017}. An analysis for differential attacks was also executed, using the Number of Pixel Change Rate (NPCR) and Unified Average Change Intensity (UACI) \cite{ozkaynak2017}. All the analyzed results showed that the developed encryption schemes have high strength and reliability, and their Mean Square Error (MSE) \cite{bani2008} is very small, which ascertains the integrity and security of the developed methods. This study has shown that the encryption method is significantly valuable in protecting the content of digital images against unauthorized access and ensuring data privacy. The contributions of our paper are as follows:
\begin{itemize}
    \item Implementation of the image encryption scheme using both 3D hyperchaotic and 2D memristive maps and comparing the effectiveness of the encryption method with the use of both schemes.
    \item Implementation of CNN in the decryption process for better accuracy and reconstruction of the original images from the encrypted ones.
    \item Comprehensive assessment of the encryption algorithms using different security measures such as entropy analysis, correlation analysis, and anomaly detection.
    \item Testing the encryption algorithm for resistance against Gaussian noise and its robustness in noisy environments.
    \item Testing the strength and robustness of the encryption scheme against possible attacks using the differential attack analysis with the help of metrics like Number of Pixel Change Rate (NPCR) and Unified Average Change Intensity (UACI).
    
    \item High accuracy of image encryption and decryption is guaranteed, proven by low Mean Squared Error (MSE) and high Structural Similarity Index Measure (SSIM) \cite{panigrahy2024} values, which means minimal deviation from the original and decrypted images.
    
    \item Empirical validation of the proposed methods over multiple test images was carried out to demonstrate the practical applicability and robustness of the encryption techniques.
\end{itemize}
The rest of the paper is organized as follows: The paper starts with the introduction summarizing the theoretical background of the novel 3D hyperchaotic and 2D memristor maps and the chaotic behaviour with their relevance to image encryption in Section \ref{section2}. Section \ref{section3} discusses the image encryption methodologies, explains the implementation of 3D hyperchaotic and 2D memristor maps, and describes the integration of CNN to enhance decryption accuracy. In Section \ref{section4}, the key generation process is discussed with a particular emphasis on the importance of strong and random keys to provide security for encryption. Section \ref{section5} elaborates on the processes for image decryption, using the CNN to reconstitute original images accurately from their encryption process. Section \ref{section6} offers a detailed security analysis based on entropy analysis, correlation analysis, anomaly detection, histogram analysis, robustness to noise, and differential attack analysis \cite{vidhya2020} to justify the robustness and effectiveness of the designed encryption algorithms. The paper concludes \ref{section7} with a summary of findings and some future research directions.

\section{Chaotic Maps} \label{section2}
The chaotic maps used here are a 3D hyperchaotic map and a 2D memristor map. These are the maps that are used as a part of the proposed algorithm.
\subsection{3D Hyperchaotic Map}
A 3D Hyperchaotic Map \cite{muni2024} is a chaotic map that describes the chaotic behaviour of a system in three dimensions. The hyperchaos in the 3D hyperchaotic map can be explained using the Lyapunov Exponent. If all three Lyapunov exponents are positive, then the system shows hyperchaotic behaviour. Such a system behaves chaotically in all three dimensions at the same time.
The equation for the 3D hyperchaotic map is as follows:

\begin{align*}
    x_{n+1} &= a_1 x_n + a_2 y_n + a_3 y_n^2 \\
    y_{n+1} &= b_1 - b_2 z_n \\
    z_{n+1} &= c x_n ,
\end{align*}

where \(x\) , \(y\) and \(z\) are the three variables over discrete time steps \(n\) and \( a_1 \), \( a_2 \), \( a_3 \), \( b_1 \),  \( b_2 \) and \(c\) are the parameters of the system. The 3D hyperchaotic map under consideration shows the prevalence of extreme hyperchaos in the sense that all three Lyapunov exponents are positive in a wide range of parameter space \cite{muni2024}.

\begin{figure}[!hbtp]
  \centering
  \includegraphics[width=0.9\textwidth]{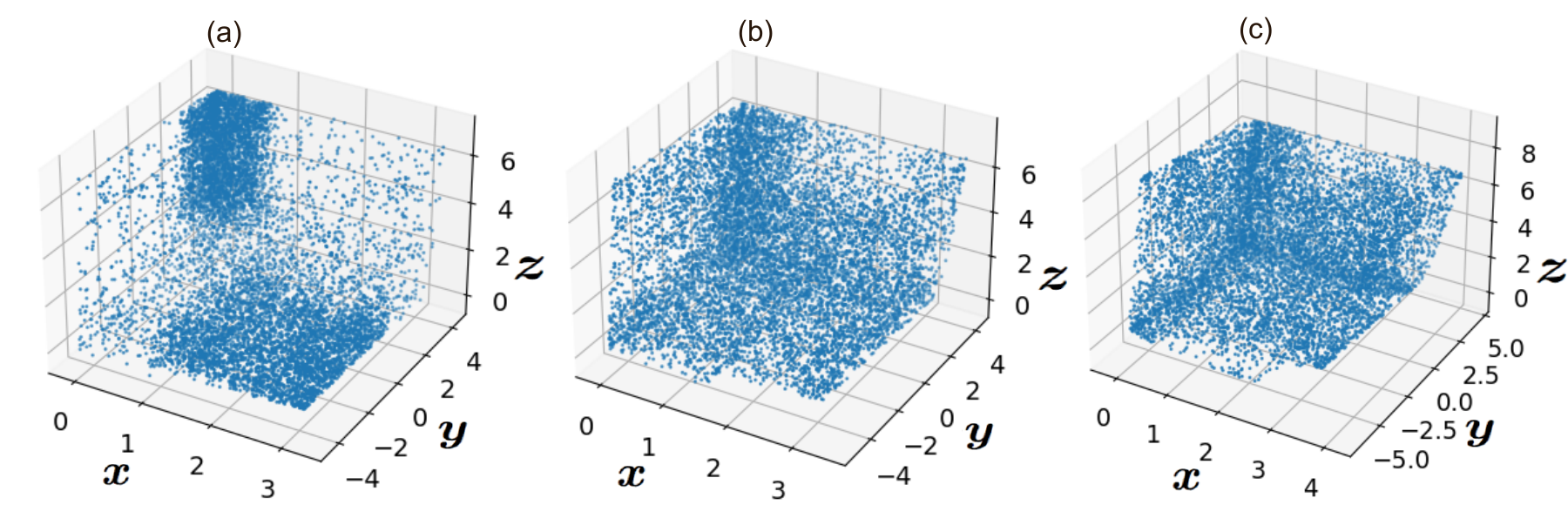}
  \caption{The figure depicts three separate three-dimensional scatter-density plots pertaining to three different trajectories in a three-dimensional hyperchaotic map; the \(x\), \(y\), and \(z\) axes display the value of the three variables involved. The trajectory in (a) is confined to such a small region of phase space that this is almost certainly the least chaotic of the three. In (b), the trajectory spreads out over a larger portion of phase space, indicating a more chaotic system than a.  And (c) is even more spread out, indicating a highly chaotic system.}
  \label{fig:3d hyperchaos}
\end{figure}

Fig\ref{fig:3d hyperchaos} illustrates the phase portrait of the hyperchaotic attractor with three positive Lyapunov exponents. In fig\ref{fig:3d hyperchaos}(a), the parameters chosen were as follows: \(a_1 = 0.03\), \(a_2 = 0.25\), \(a_3 = 0.11\), \(b_1 = 4\), \(b_2 = 1.2\) and \(c = 2.15\) and it shows the hyperchaotic cube like attractor. By changing \(a_1\) to 0.05 and keeping the rest of the parameters as before, a small change in \(a_1\) results in a very significant difference in the system's behaviour, showing dynamic instability in \ref{fig:3d hyperchaos}(b). For the fig\ref{fig:3d hyperchaos}(c), the values chosen were as follows: \(a_1 = 0.09\), \(a_2 = 0.30\), \(a_3 = 0.12\), \(b_1 = 4\), \(b_2 = 1.2\), and \(c = 2.15\). Out of all three, the second configuration was chosen to be used as the input for the proposed encryption method.

\begin{figure}[!hbtp]
  \centering
  \includegraphics[width=0.4\textwidth]{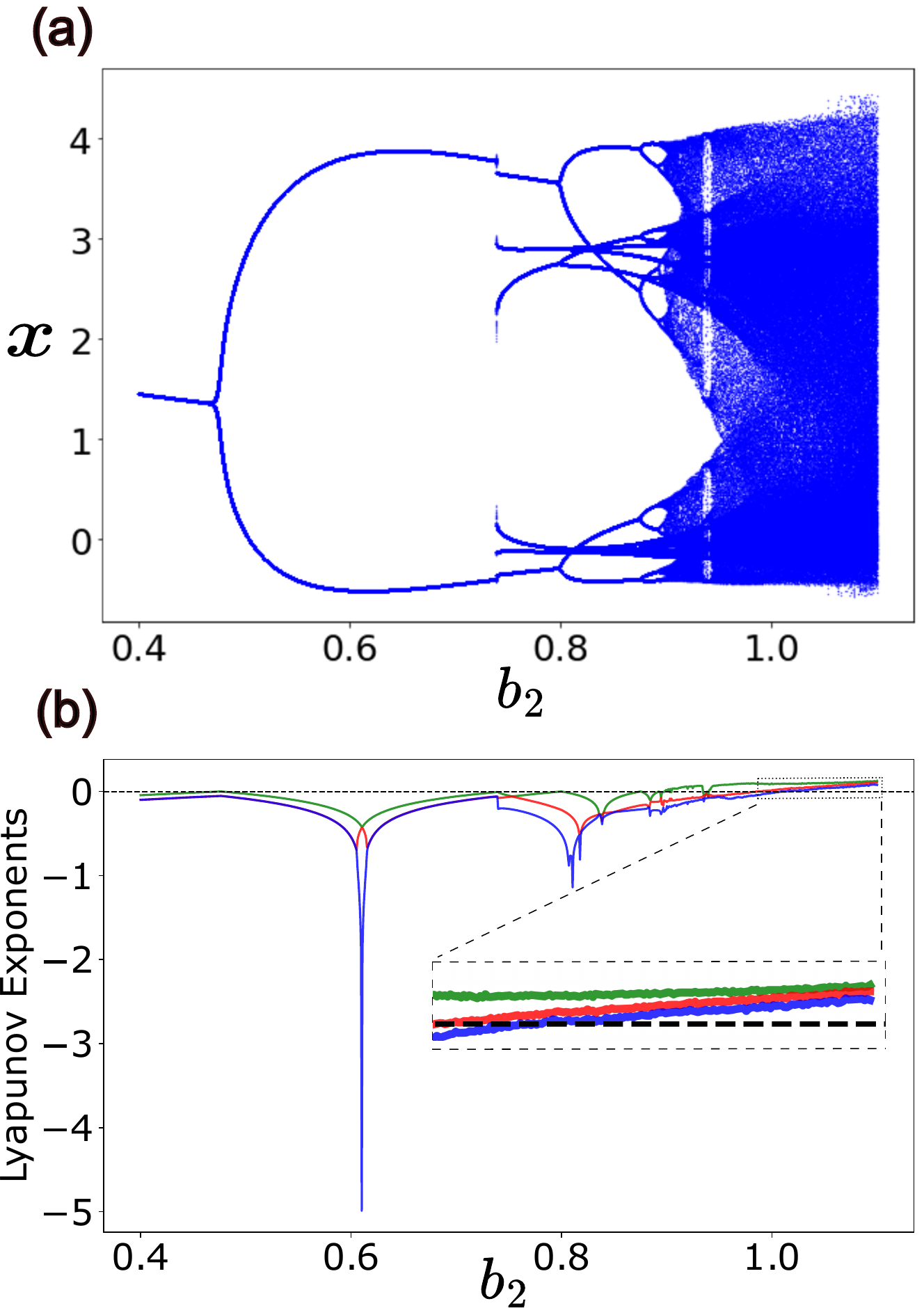}
  \caption{Bifurcation diagram and Lyapunov exponent spectrum of 3D hyperchaotic map is given\cite{muni2024}. It shows that this system is showing very complex hyperchaotic behaviour as the parameter \(b_2\) of the system changes its values. Figure (a) represents the Bifurcation diagram for a 3D hyperchaotic map. Figure (b) Lyapunov exponent diagram for the above 3D hyperchaotic map clearly indicates the stability and chaotic region as a function of changing parameters.}
  \label{bif2}
\end{figure}
A one-parameter bifurcation diagram of \(x\) vs \(b_2\) along with the corresponding Lyapunov exponent spectrum is shown in figure:\ref{bif2}. A period-doubling cascade which occurs as \(b_2\) increases eventually gives way to chaotic behaviour at higher values of \(b_2\), as evident by the positive Lyapunov exponents implying the chaotic nature of the system with further increase in \(b_2\), the system transits to extreme hyperchaotic behaviour with three Lyapunov exponents positive. (see figure:\ref{bif2} [b]).

The next section deals with the novel 2D quadratic memristor map which also exhibits hyperchaotic behaviour.

\subsection{2D Memristor Map}
Memristors \cite{ye2020}are nonlinear electrical elements whose resistance is based on the past electric charge flow history. In the 2D memristor map, typically, two variables depict the state of the system while these evolve over time according to some dynamical equations defined in two dimensions.
The equation of a 2D Memristor map is defined as follows:
\begin{align}
    x_{n+1} &= k((q_n^2) - 1) x_n \\
    q_{n+1} &= q_n + x_n ,
\end{align}
where $x_n$ and $q_n$ are the state variables and \(k\) represents the parameter.

\begin{figure}[!hbtp]
  \centering
  \includegraphics[width=0.9\textwidth]{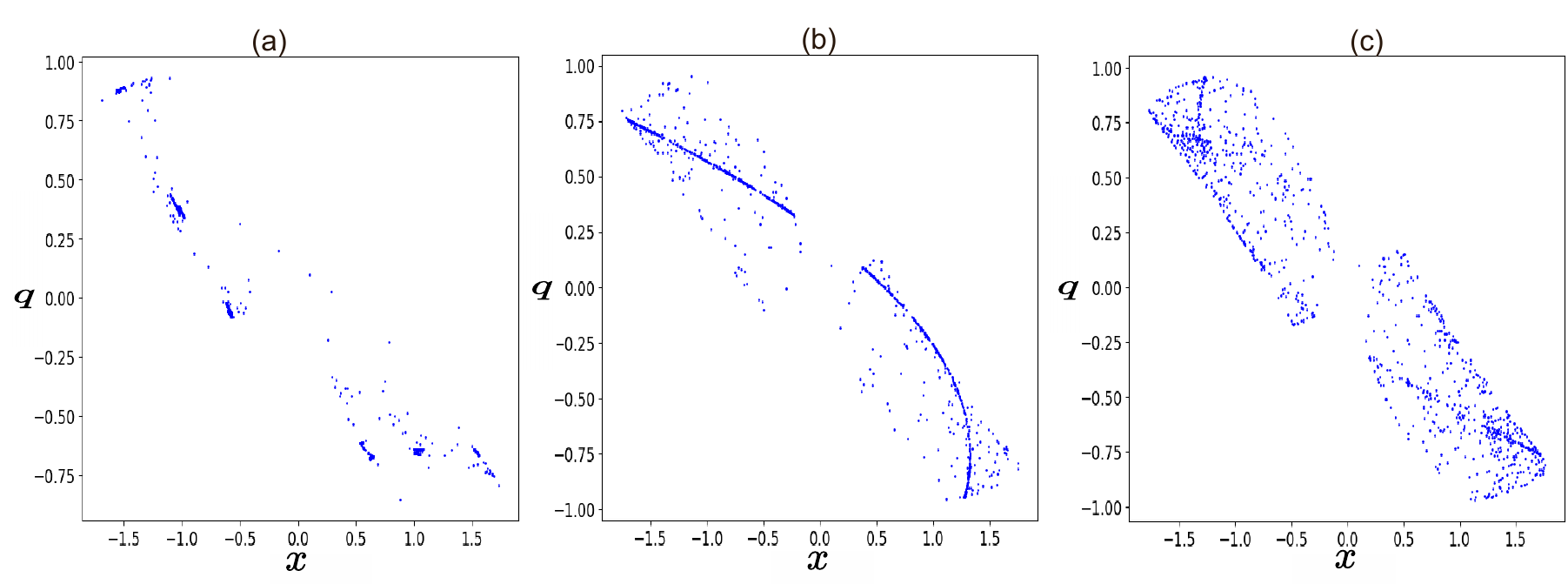}
  \caption{Evolution of hyperchaotic behaviour of 2D memristor map is shown. Fig(a) here illustrates the phase space at \(k = 1.74\), which is a hyperchaotic regime. Fig(b) here illustrates the phase space at \(k = 1.75\), which lies in a hyperchaotic regime. Fig(c) here illustrates the phase space at \(k = 1.76\), which lies in two disjoint cyclic hyperchaotic regimes.}
  \label{2d mem}
\end{figure}

Memristors are special resistors because the resistance value through them is based on history. This particular memory-dependent behaviour is represented in the 2D memristor map dynamics and forms part of its unique features: hysteresis and non-volatility. The evolution of the hyperchaotic behaviour of the map is shown in figure \ref{2d mem} via the phase portraits in \(x\)-\(q\) plane in (a) chaotic/hyperchaotic attractor is shown with Lyapunov exponents (0.0527,-0.0597). In (b), disjoint cyclic hyperchaotic attractors are shown with Lyapunov exponents(0.2368,0.0263). In (c), a hyperchaotic attractor with much mixing is shown with Lyapunov exponents(0.2370,0.1009) at \(k=1.76\) is shown.

\begin{figure}[!hbtp]
  \centering
  \includegraphics[width=0.5\textwidth]{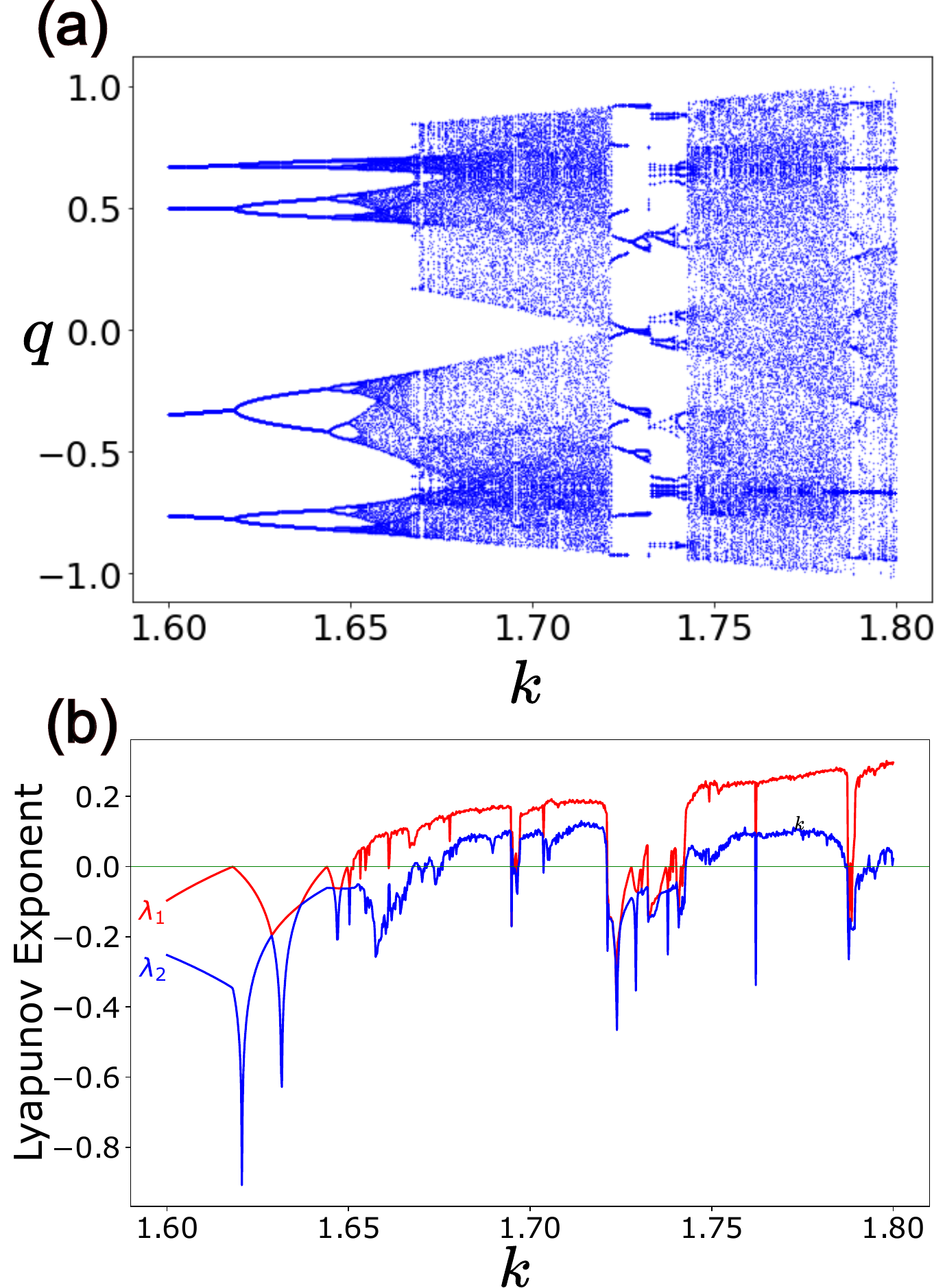}
  \caption{The bifurcation diagram and Lyapunov exponent spectrum of the 2D memristor map are shown. (a) shows the bifurcation diagram showing the state of the system developing as it depends on the parameter \(k\); (b) shows the Lyapunov exponent spectrum showing the existence of hyperchaos depending on the parameter \(k\).}
  \label{bif1}
\end{figure}

In Figure:\ref{bif1}, both the bifurcation diagram and the Lyapunov exponent diagram of the 2D memristor map show the prevalence of hyperchaos over a wide range of parametric space. The system has a stable fixed point when the parameter \(k\) is small. This is evidenced by the single branch in the bifurcation diagram and both negative Lyapunov exponents. At \(k=1.76\), the map discrete stable period four-orbit, and as further \(k\) increases, it undergoes a period-doubling bifurcation chaos and subsequently to hyperchaos.

\section{Encrypting the images using discrete hyperchaotic maps } \label{section3}
\subsection{Using 3D Hyperchaotic map}
In this paper, we show a novel method of image encryption using a hyperchaotic map \cite{gao2022} along with a hybrid neural network technique for decrypting images, which we have termed a Convolutional Neural Network (CNN).  The basics of the encryption algorithm depend on the hyperchaotic map that surges intricate routes of chaos. Therefore, they scramble for image pixels to keep safety at stake. For example, the hyperchaotic map is defined provided that nonlinear iterative maps have initial values set and parameters manipulated to generate the volatile sequences. Initially, we obtain these sequences by starting the variables and moving with successive applications of the hyperchaotic map equations. These give rise to three chaotic sequences: \(x\), \(y\), and \(z\), each of which is instrumental in the encryption process. To encrypt the image, we convert it into a \texttt{NumPy} array. Then, it encrypts every photograph pixel through a number set operations using chaotic sequences such as bitwise XOR and modular arithmetic. This procedure would efficiently shuffle the pixel values and thus make the image completely unrecognizable if chaotic sequences in place are not employed as the right ones. The formula used for the above encryption is given below:
\begin{enumerate}
    \item \textbf{XOR Operation}:
        \[ E_{i,j} = P_{i,j} \oplus \left\lfloor X_{i,j} \cdot 255 \right\rfloor \]
        
    \item \textbf{Addition Operation}:
        \[ E_{i,j} = (P_{i,j} + \left\lfloor Y_{i,j} \cdot 255 \right\rfloor) \mod 256 \]
        
    \item \textbf{Subtraction Operation}:
        \[ E_{i,j} = (P_{i,j} - \left\lfloor Z_{i,j} \cdot 255 \right\rfloor) \mod 256 \]
    \label{operations}    
\end{enumerate}
In  steps 1-3, $P_{i,j}$ denotes the pixel value at position \((i,j)\) of the original image, $E_{i,j}$ denotes the encrypted pixel value at position \((i,j)\) of the encrypted image,
$X_{i,j}$, $Y_{i,j}$, $Z_{i,j}$ represent the corresponding values from the  sequences derived from the\(x\), \(y\) and \(z\) sequence of the hyperchaotic map respectively. $\oplus$ denotes the bitwise XOR operation. $\lfloor \cdot \rfloor$ denotes the floor function, which rounds down to the nearest integer
\begin{figure}[!hbtp]
  \centering
  \includegraphics[width=0.7\textwidth]{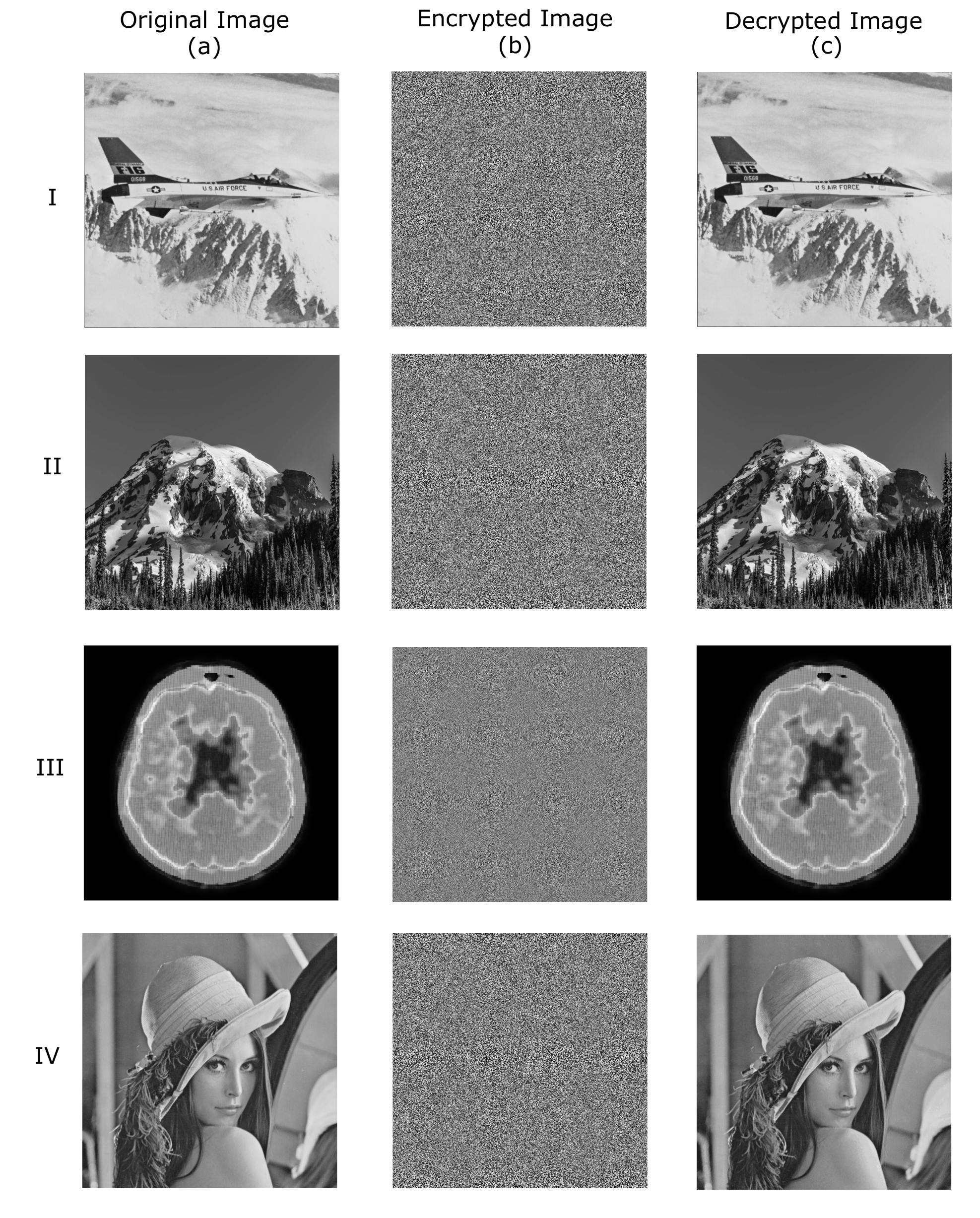}
  \caption{An original greyscale image (a) is encrypted by a 3D hyperchaotic equation and converted into an unreadable, scrambled image (b). A backward process leads to the decryption of the picture and its return to its original format (c). Therefore, it is possible to examine how efficiently a technique encrypts or decrypts a picture.}
  \label{fig:3}
\end{figure}

\newpage
\subsection{Using 2D Memristor Map}
We propose a novel technique for image encryption and decryption \cite{qian2023} by incorporating a 2D quadratic memeristive hyperchaos map with a convolution neural network (CNN). The encryption algorithm depends on a chaotic map generating complex sequences for scrambling pixel values and, therefore, increasing security. A chaotic map refers to a set of nonlinear equations which initiate series from particular values and parameters. In particular, we utilize the chaotic map formula that needs initial conditions \(x\), \(q\) alongside a given parameter \(k\) to yield fresh outputs step by step. So as to get these series, it is necessary to start with certain \(x\) and \(q\) values, including running chaotic map iterations over the number of pixel times, thereby producing adequate sequence for the encryption method. The image is first loaded and translated into a \texttt{NumPy} array to start the encryption procedure. Each pixel is encrypted using generated chaotic sequences after resizing the image to appropriate pixels and converting it into an RGB format. This means manipulating the pixel values and conserving the correct order through a series of operations with chaotic sequences like bitwise XOR and modular arithmetic. The encrypted image thus looks like random noise, rendering it meaningless without the proper chaotic sequences. The equation for encryption is given below:

\[
\text{encrypted\_pixel} = \text{pixel} \oplus \lfloor x_{\text{sequence}[index]} \times 255 \rfloor
\]

\[
\text{encrypted\_pixel} = \text{encrypted\_pixel} + \lfloor q_{\text{sequence}[index]} \times 255 \rfloor
\]

\[
\text{encrypted\_pixels}[i, j, k] = \text{encrypted\_pixel} \mod 256 ,
\]
where:
\begin{itemize}
    \item \(encrypted\_pixel\) represents the value of the encrypted pixel.
    \item \(pixel\) denotes the original pixel value.
    \item \(x_{\text{sequence}[index]}\) is the \(x\) sequence value corresponding to the current index.
\item \({encrypted\_pixel}\) represents the updated value of the encrypted pixel after addition.
    \item \(q_{\text{sequence}[index]}\) is the \(q\) sequence value corresponding to the current index.
\item \({encrypted\_pixels}[i, j, k]\) denotes the value of the encrypted pixel in the \(i\)-th row, \(j\)-th column, and \(k\)-th channel.
\end{itemize}
\begin{figure}[!hbtp]
  \centering
  \includegraphics[width=0.8\textwidth]{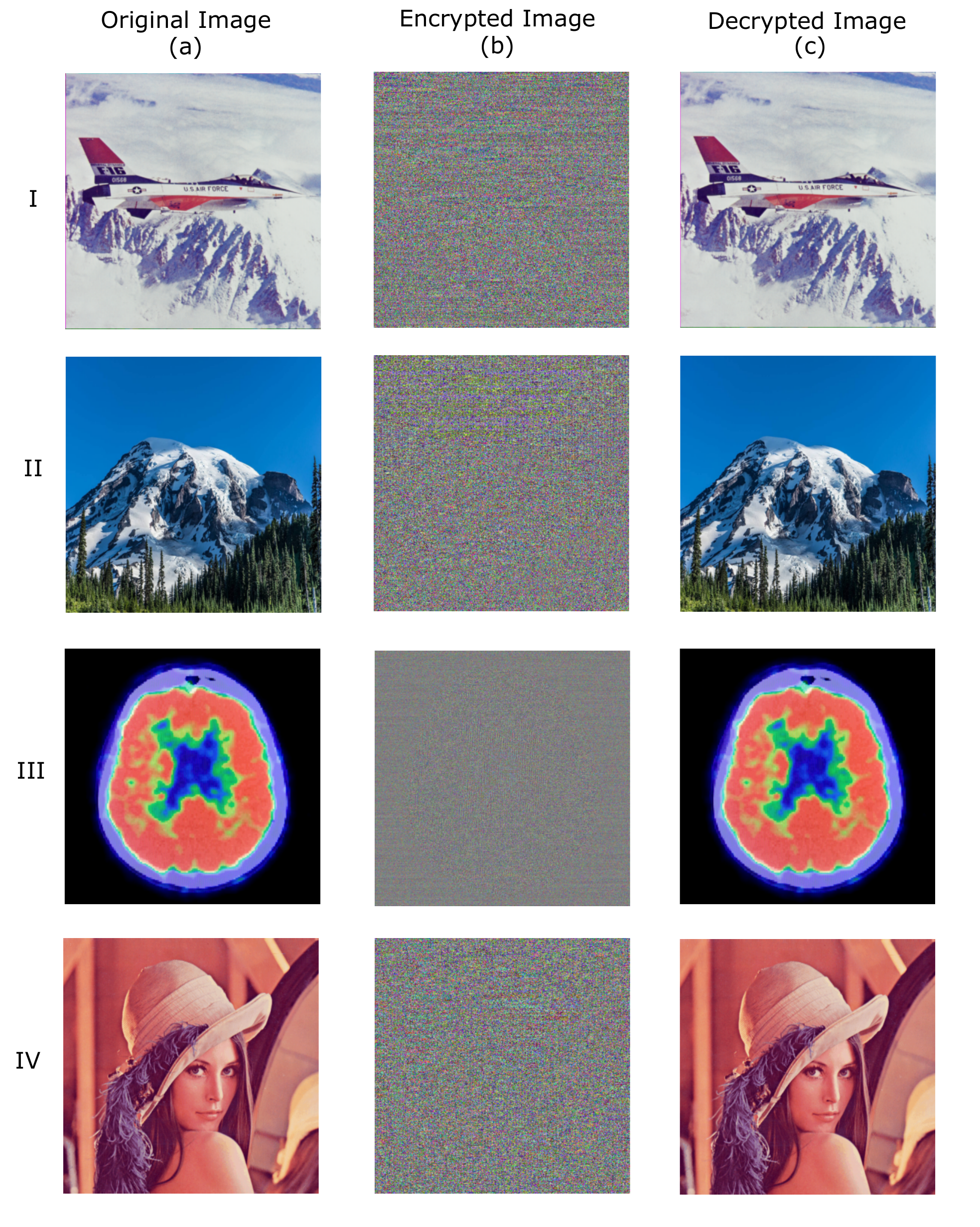}
  \caption{ The effectiveness of encryption and decryption of four colour images by two-dimensional memristor map is shown. The images are organized in rows, where each row presents the original image (a), its corresponding encrypted image (b), and the decrypted image (c). The four sets of distinct images specify how the encryption process is reliable. Reliable encryption and successful decryption ensure security and privacy while sending and storing sensitive information.}
  \label{fig:4}
\end{figure}

\newpage
\section{Key Generation} \label{section4}
The strong key generated in the encryption process of an image is very important as it adds security and guarantees privacy \cite{ding2021}. The generated key is highly confidential and acts as a basis for determining how every pixel of the picture changes during encoding. Randomness and complexity are added to the method by which an algorithm scrambles data, thus making it impossible computationally for unauthenticated individuals to discover or decrypt any encrypted picture unless they have the correct keys. Therefore, by making powerful, unique keys, we ensure that our systems are well-guarded against unauthorized entry while still keeping safe contents and the integrity and confidentiality of images intact. Size and randomness also contribute to its strength, thereby enhancing resistance against brute force attacks and other cryptographic weaknesses. Key generation, on the other hand, means the process followed to derive a sequence of characters probabilistically that forms the basis of data encryption and decryption. This will use the secrets module to gain access to random sources, using a strong source of secure randomness provided by the operating system. This factor will make the generated key very unpredictable and ensure high resistance toward cryptographic attacks. Here, for this algorithm, the module generates a random hexadecimal string of the length specified for generating a strong cryptographic key. The key also determines a huge part of the process in how every image pixel will be transformed during encoding into conditions of assuring the created image data is both confidential and integral. In addition, the randomization and complexity of the generated key add to the overall strength of the encryption system to avoid entry by unauthorized persons and avert any vulnerability in its cryptographic system towards improved security.

\section{Decrypting the images using discrete hyperchaotic map} \label{section5}
\subsection{Using 3D Hyperchaotic Map}
Decryption is fundamentally the opposite of encryption. It utilizes identical disordered sequences for unscrambling pixels so that they are returned to their original state as an image. It serves as an essential stage in ensuring that the initial image is obtainable from encrypted information and that any encryption program can be rolled back. For the purpose of improvement, a CNN has been used to evaluate the encrypted image. The CNN comprises different convolutional, pooling, and upsampling layers. Its form has been chosen to detect complicated patterns in images. Training of the CNN involves the process of making the encrypted image map to what it looked like before. This will encompass changing the image's shape to fit into the network requirements and normalising data. The CNN is built using the Adam optimizer \cite{alsafyani2023} and a mean squared error (MSE) \cite{ahmad2010} loss function, which are standard choices for training these models. The weights are adjusted with each iteration through training in order to decrease loss functions. The complete process can be seen when showing the original, encrypted, and decrypted images on one screen. It provides a point of comparison, making evident the efficiency of the encryption and decryption practices. The encrypted version of an image looks like a random scrambling, signifying secureness in the process used, but the decrypted image almost resembles the original image, meaning that the entire scenario was correct.
The decryption is done as follows:
\begin{itemize}
    \item  An array of zeros is initialized where pixels of the decrypted image will be stored. Make sure that this has the same shape as the encrypted image.
    \item find the width and height of the encrypted image. Among these, compute the total steps, which are the product of width times height. 
    \item truncate chaotic sequences according to the total steps.
    \item for every pixel, compute its linear index with respect to its row and column position.
    \item the value obtained from the corresponding position in the \(z\) sequence to the value of the encrypted pixel and take mod 256.
    \item Subtract the value obtained from the corresponding position in the \(y\) sequence from the result and take mod 256.
    \item It is then XORed with the value obtained from the corresponding position in sequence \(x\). It is taken mod 256.
    \item Store the final decrypted pixel value into the corresponding position of the decrypted pixel array

\end{itemize}

\begin{figure}[!hbtp]
  \centering
  \includegraphics[width=1.0\textwidth]{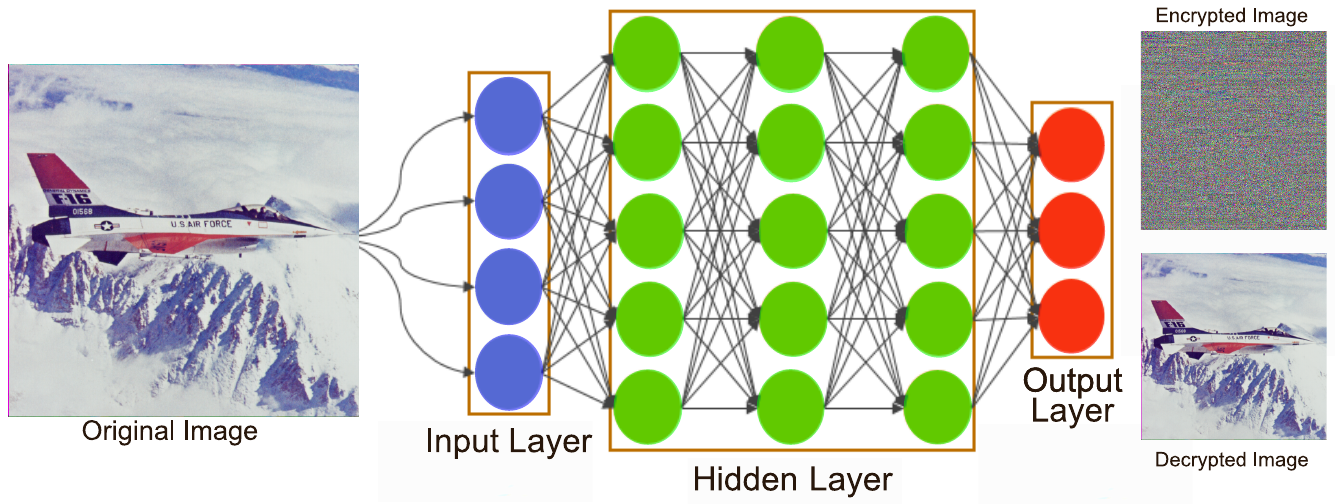}
  \caption{The figure explains the working of a neural network for image encryption and decryption. The original image is put in the input layer, and the information in the hidden layer is processed. Later, the output layer gives the encrypted image, and this process can be reversed to decrypt the image.}
  \label{fig:1}
\end{figure}

\subsection{Using 2D memristor map}

Decryption ensures that an encryption algorithm is reversible, and the original image can be restored perfectly from encrypted data by using the same chaotic sequences in the inverse method to take pixels back to their initial values. 
\begin{itemize}
    \item Modular arithmetic regains the decrypted pixels with a bitwise XOR operation.
    \item CNN analyzes and reconstructs the encrypted image.
    \item Convolutional, pooling, and upsampling layers constitute the architecture of CNN and comprehend intricate patterns within images.
    \item After training, the CNN maps the encrypted image to its original form.
    \item In this step, the data normalization and shape adjustment are done to match the network input requirements.
    \item Optimization uses the Adam optimizer and Mean Squared Error (MSE).
    \item Models get trained over 10 epochs to minimize errors and enable them to converge to real images.
    \item Both the images encrypted and decrypted are random sets of noise. Moreover, clear images that turn out are decrypted upon matching the original.
\end{itemize}

\begin{figure}[!hbtp]
  \centering
  \includegraphics[width=1\textwidth]{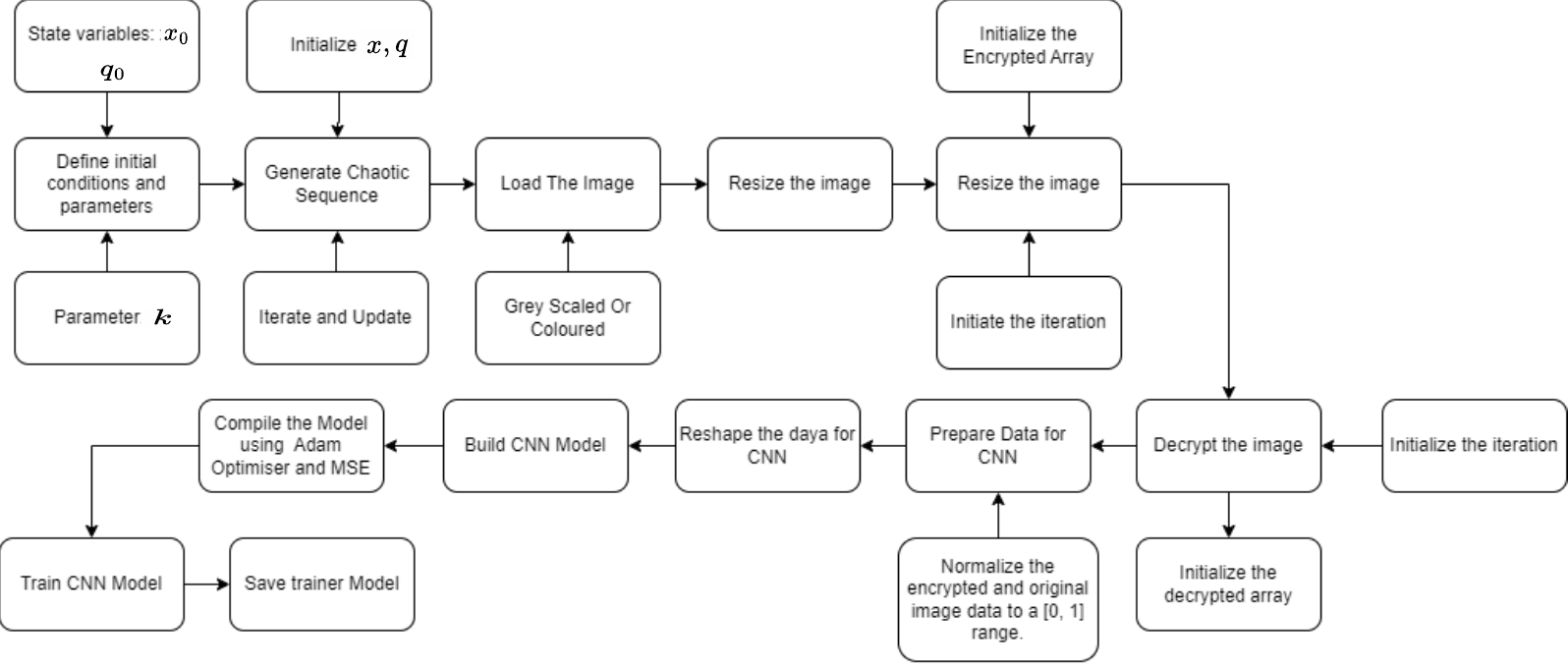}
  \caption{The provided flowchart presents the entire process of utilizing hyperchaotic sequences and Convolutional Neural Network (CNN) in encrypting and decrypting an image. This outline involves initializing parameters, creating chaotic sequences and processing images, and training CNN models, resulting in image decryption in the end, thereby assuring confidentiality and integrity during both the encryption and decryption stages.}
  \label{fig:5}
\end{figure}

\newpage

\section{Security Analysis} \label{section6}
The evaluation of image encryption and security is a measure of how effective and strong encryption algorithms are for the purpose of protecting image data from unauthorised access and tampering. This analysis incorporates several features. Initially, it ensures confidentiality through hiding, whereas the plain image content is hidden from unauthorised parties. This involves evaluating the resistance of encryption used against attempts at unauthorized decryption without the key for decryption. Next is the concept of integrity: the encrypted image shall remain unchanged during its transmission or storage and be checked against modification. Authenticity verification defends the integrity of the image and assures that no fraud is being carried out with strong encryption techniques. The next key aspect is resiliency against cryptographic attacks, from the point of standing against the various attack vectors and ensuring that the image is secure. For the effective management of encryption keys, it is required to securely generate, store, and distribute keys, and thereby test their strength and the efficiency of the key exchange. Finally, the computational complexity has to be considered to ensure efficient encryption and decryption processes since these have to be carried out at different types of computing base units while keeping the security levels satisfactory to ensure greater flexibility of the encryption system.

\subsection{Entropy analysis} 
Entropy is the statistical measure of randomness or uncertainty in the image data. The image is flattened into a one-dimensional array for facilitating pixel-level analysis\cite{hausen2020}. The pixel distribution is calculated by counting the occurrence of each pixel value in the flattened image. In the next step, the probability of the pixel value is calculated by dividing the pixel distribution by the net number of image pixels. In the end, the entropy is found by using the equation \eqref{eq:entropy}. The resulting entropy value gives the level of uncertainty in the encrypted image. The output value of entropy is found to be 7.598 approximately (refer Table:\ref{entropy analysis} and Table:\ref{entropy grey}). This indicates a moderately high complexity level in the encrypted image.

\begin{equation}
H(X) = -\sum_{i=1}^{n} P(x_i) \log_2(P(x_i))
\label{eq:entropy}
\end{equation}

Where:
\begin{itemize}
    \item \( P(x_i) \) is the probability of occurrence of the \( i \)th value in the dataset.
    \item \( n \) is the total number of unique values in the dataset.
    \item \( \log_2 \) represents the logarithm to the base 2.
\end{itemize}

\begin{table}[!hbtp]
\centering
\begin{tabular}{cccc}
\toprule
\toprule
\multicolumn{1}{c}{\textbf{Color Image}} & \multicolumn{3}{c}{\textbf{Entropy}}\\
\cmidrule(rl){2-4} 
\textbf {} & {Original} & {Encrypted } & {Decrypted} \\
\midrule
Aeroplane & 6.57683 & 7.97806 & 6.57683 \\
Brain & 5.38622 & 7.96943 & 5.33862 \\
Lena & 5.10141 & 7.99174 &  5.10141  \\
Rainier & 6.76169 & 7.98567 &  6.76169  \\
\bottomrule
\bottomrule
\end{tabular}
\caption{Color image entropy analysis. The values of entropy for the original, encrypted, and decrypted images are shown in. A higher value of entropy indicates increased randomness; hence, for the encrypted images, security is improved.}
\label{entropy analysis}
\end{table}

\begin{table}[!hbtp]
\centering
\begin{tabular}{cccc}
\toprule
\toprule
\multicolumn{1}{c}{\textbf{Grey Scaled Image}} & \multicolumn{3}{c}{\textbf{Entropy}}\\
\cmidrule(rl){2-4} 
\textbf {} & {Original} & {Encrypted } & {Decrypted} \\
\midrule
Aeroplane & 6.70245& 7.53609 & 6.70245 \\
Brain & 5.55853 & 7.99991 & 5.55853 \\
Lena & 7.44550 & 7.99934 &  6.87077  \\
Rainier & 6.98145 & 7.63111 &  6.98145  \\
\bottomrule
\bottomrule
\end{tabular}
\caption{The entropy of the original, encrypted, and decrypted grayscale images. A high entropy value indicates a more random distribution of pixels in an image.}
\label{entropy grey}
\end{table}

\newpage 
\subsection{Correlation Analysis}
Correlation is a statistical measure that explains the degree of linear relationship present between the neighbouring pixels of an encrypted image. It also measures how much one pixel's intensity value is related to another one. Positive coefficients between an original colour channel and a decrypted one provide assurance that the method preserves the integrity of an image. Negative coefficients for channels encrypted before attested to its successful encryption; positive coefficients for the channels after attested to the effective restoration of the image. Similar contents close to 1.0 show good encryption and decryption, while those close to 0 will show the image as being well-encrypted and difficult to recover. This confirms that the method used here is efficient for encryption and decryption Table:\ref{correlation}.

\[
\rho_{X,Y} = \frac{\sum_{i=1}^{n}(X_i - \bar{X})(Y_i - \bar{Y})}{\sqrt{\sum_{i=1}^{n}(X_i - \bar{X})^2} \sqrt{\sum_{i=1}^{n}(Y_i - \bar{Y})^2}}
\]

Where:
\begin{itemize}
    \item \( \rho_{X,Y} \) is the Pearson correlation coefficient between variables \( X \) and \( Y \).
    \item \( X_i \) and \( Y_i \) are the individual data points in the datasets \( X \) and \( Y \), respectively.
    \item \( \bar{X} \) and \( \bar{Y} \) are the mean values of datasets \( X \) and \( Y \), respectively.
    \item \( n \) is the total number of data points.
\end{itemize}

\begin{table}[htbp]
    \centering
    \captionsetup{position=bottom} % Caption at the bottom
    
    \begin{tabular}{cccccc}
        \toprule
        \multicolumn{5}{c}{\textbf{Correlation Coefficients for Coloured Images}} \\
        \midrule
        \textbf{Image} & \textbf{Channel} & \textbf{Original} & \textbf{Encrypted} & \textbf{Decrypted} \\
        \midrule
        Aeroplane & Blue & 0.963 & -0.119 & 0.963 \\
        & Green & 0.942 & -0.103 & 0.942 \\
        & Red & 0.973 & -0.093 & 0.973 \\
        \midrule
        Brain & Blue & 0.951 & -0.139 & 0.951 \\
        & Green & 0.934 & -0.139 & 0.934 \\
        & Red & 0.968 & -0.153 & 0.968 \\
        \midrule
        Lena & Blue & 0.931 & -0.077 & 0.931 \\
        & Green & 0.964 & -0.100 & 0.964 \\
        & Red & 0.973 & -0.093 & 0.973 \\
        \midrule
        Rainier & Blue & 0.957 & -0.052 & 0.957 \\
        & Green & 0.924 & -0.086 & 0.924 \\
        & Red & 0.949 & -0.158 & 0.949 \\
        \bottomrule
    \end{tabular}

    \caption{Correlation coefficients for blue, green, and red channels of four different images in original, encrypted, and decrypted states: "Aeroplane", "Brain", "Lena", and "Rainier". Positive correlation coefficients are obtained for the original and decrypted channels, indicating successful decryption. In the case of encrypted channels, all the coefficients were negative, revealing a very significant alteration during encryption. These results clearly suggest the effectiveness of the encryption process in conserving the original information of the images while allowing successful decryption.}
    \label{correlation}
\end{table}

\begin{figure}[!hbtp]
  \centering
  \includegraphics[width=1.0\textwidth]{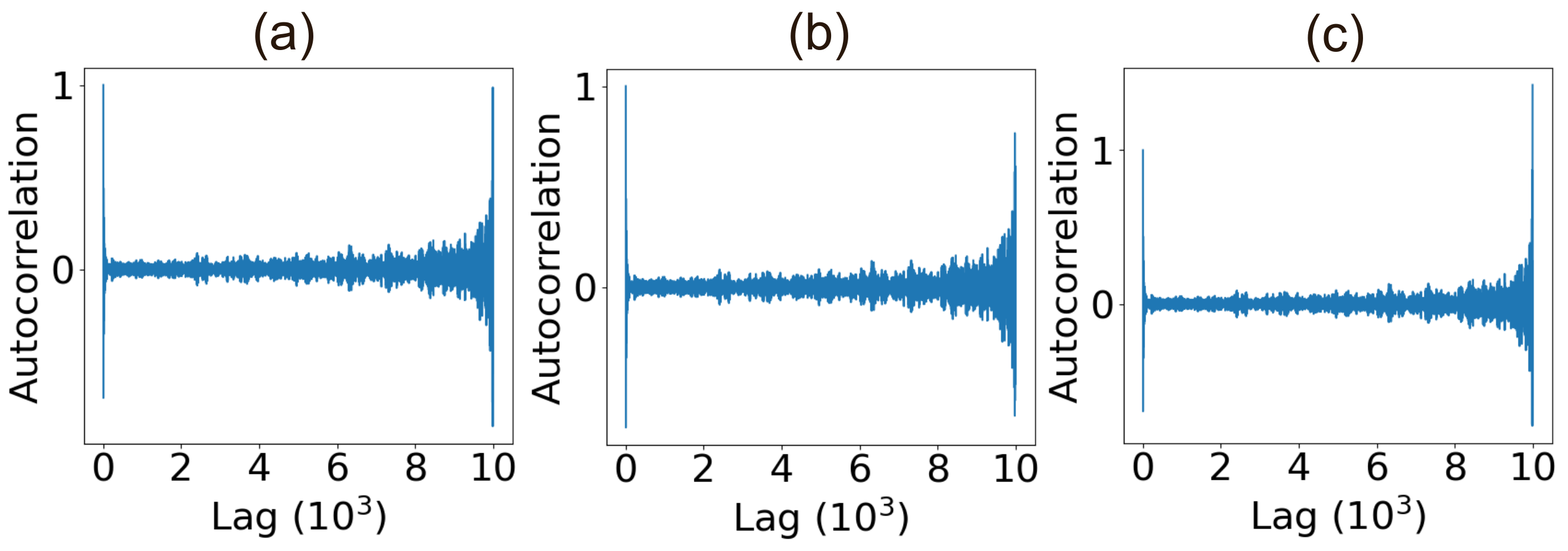}
  \caption{The autocorrelation plots for the variables \(x\), \(y\), and \(z\) sequences of the 3D hyperchaotic map are shown in (a), (b), and (c), respectively. These sequences describe the correlation of any sequence with itself while shifting the time lag. The insufficiency of significant autocorrelation in these plots indicates that each sequence should exhibit chaotic and unpredictable behaviour, which is not dependent on the memory of past values. It means, then, that such a lack of auto-correlation becomes its intrinsic property, which confirms the existence of chaoticity in the 3D hyperchaotic map.}
  \label{autocorrelation}
\end{figure}

\newpage
\subsection{Anomaly Detection}
Anomaly detection, also known as outlier detection, is applied in many areas to find natural patterns or cases that differ significantly from a general rule or context. In the context of image encryption and decryption, the importance of an anomaly detection method basically lies in the evaluation of the effectiveness of encryption algorithms in maintaining image quality and integrity. Anomaly detection is applied to compare decrypted images with their original ones and can find any anomaly that might be introduced during the encryption and decryption process. In our study, we applied anomaly detection techniques for the performance evaluation of our encryption algorithm. Measures like Mean Squared Error and Structural Similarity Index were applied to detect the differences between original and decrypted images. The resultant anomaly detection analysis was then tabulated, previewing MSE and SSIM values for each image (Aeroplane, Brain, Lena, and Rainier) tested in our study. The task of performing anomaly detection was done in a series of processes. First, we encrypted the original images using our encryption algorithm. Second, the encrypted images were decrypted to get the reconstructed images. Third, we compared the reconstructed image with the original images using MSE and SSIM metrics. The low MSE value will show that the reconstructed image shows a low difference between the original and the decrypted images, and a higher SSIM value will show that the images have a higher similarity. Anomaly detection revealed practical value in the performance results of our encryption and decryption algorithm. For instance, mostly high SSIM values and low MSE values would indicate that our algorithm was able to preserve the quality and integrity of images from encryption to decryption. However, any kind of anomalies or deviations detected by anomaly detection pointed out the grey areas where the encryption algorithm or decryption procedure should be improved.

\begin{table}[htbp]
    \centering
    \begin{tabular}{|c|c|c|}
        \hline
        \textbf{Coloured Images} & \textbf{MSE} & \textbf{SSIM} \\
        \hline
        Aeroplane & $1.27 \times 10^{-6}$ & 0.9999999999647077 \\
        Brain & $1.51 \times 10^{-7}$ & 0.9999999999370367 \\
        Lena & $1.27 \times 10^{-6}$ & 0.9999999995838884 \\
        Rainier & $1.61 \times 10^{-7}$ & 0.9999999999330903 \\
        \hline
    \end{tabular}
    \caption{Anomaly Detection performance for Coloured Images. The Mean Squared Error (MSE) and Structural Similarity Index (SSIM) values show the accuracy of the anomaly detection algorithms applied to the Aeroplane, Brain, Lena, and Rainier images.}
    \label{mse ssim table}

\end{table}
The MSE values are represented in Table \ref{mse ssim table} and give the average squared difference between the original and decrypted images. The smaller the MSE values, the smaller the discrepancy between the two images; in other words, the higher the fidelity in preserving the details of an image in encryption and decryption. SSIM values fall between 0 and 1, with a perfect similarity of 1 between the original and decrypted images. According to Table \ref{mse ssim table}, the MSE values for all the images are very small, showing very minimal changes between the original and decrypted images. The SSIM values are almost 1, indicating a high resemblance. The results indicate that the process of encryption and then decryption both retained the quality and integrity of the images. The low MSE values show that the encryption algorithm kept the main features of the images, while high SSIM values confirm that the decrypted images do look a lot like the original images.

The results of anomaly detection definitely show that the process of image encryption and decryption is robust. The very small differences between the original and the decrypted images, shown in the MSE and SSIM results, confirm that the encryption algorithm can well keep image quality and integrity. This supports the fact that the process of encryption and decryption used in the study is reliable and robust in ensuring the security of data and integrity in the process of transmitting and storing images.

\subsection{Histogram Analysis}
The histogram analysis is one of the most applied methods in image processing to research the intensity distribution of pixels across the image \cite{kamal2021}. It highlights all important information about contrast, brightness, and the general arrangement of the image by plotting the number of occurrences of each of the intensity values versus the histogram graph. Hence, histogram analysis will be integral in examining the integrity of an image and the effectiveness and reliability of the encryption process. The effect of encryption and decryption on pixel intensity distribution was examined using a histogram. By analyzing the histogram of the original images and their corresponding encrypted and decrypted ones, the study was able to draw conclusions about the distortion and safeguarding of image features through the process of encryption and decryption.

\begin{figure}[!hbtp]
  \centering
  \includegraphics[width=1\textwidth]{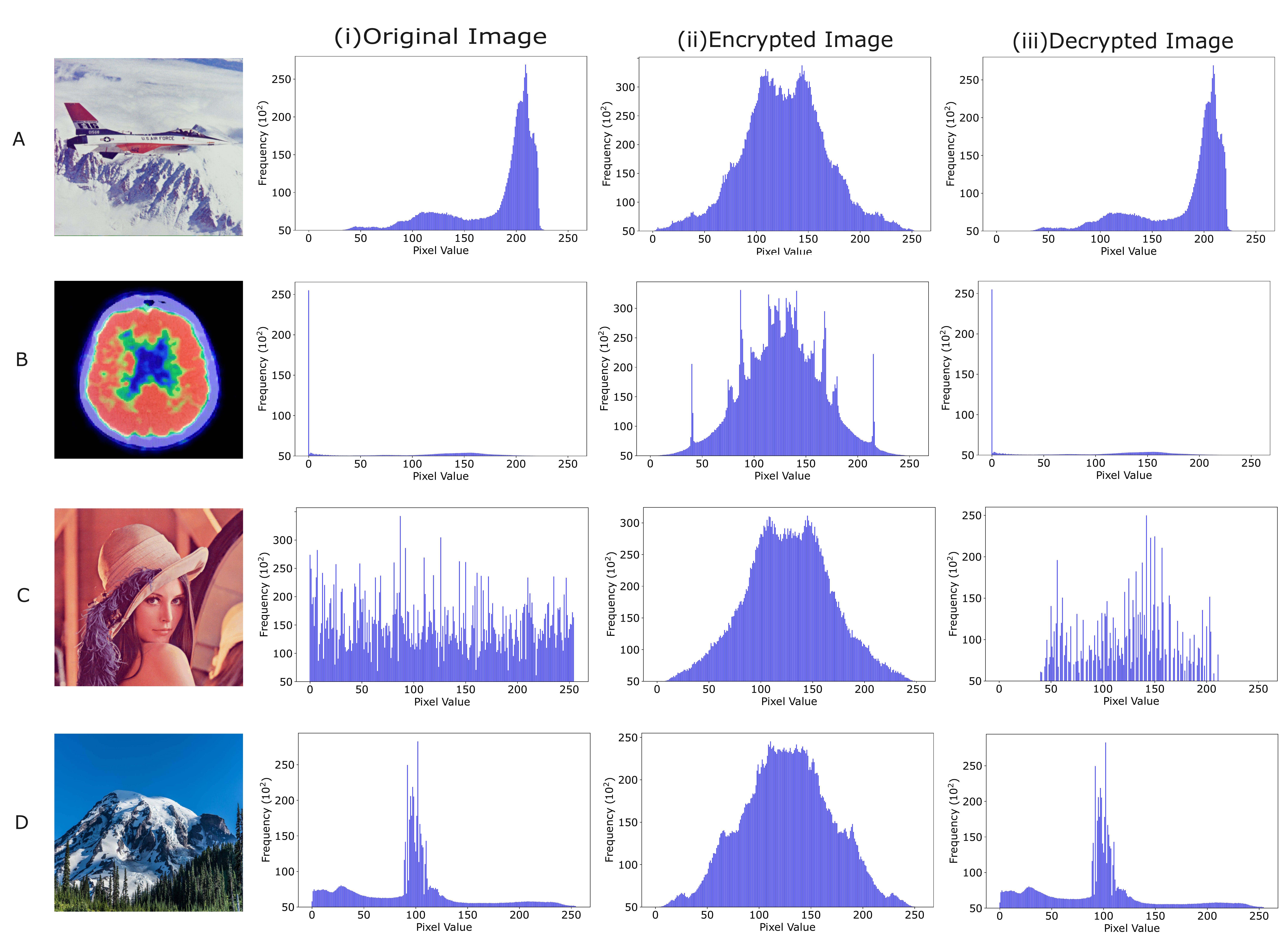}
  \caption{Effectiveness of encryption and decryption method for different images, namely A, B, C, and D, are shown. The original images (i) are encrypted for a uniform pixel value distribution (ii). Upon decryption, the original pixel value distributions are restored (iii), indicating the robustness of the process of encryption and decryption in securing sensitive image information.}
  \label{histogramanalysis}
\end{figure}
\newpage
\subsection{Robustness to Noise Analysis}
\begin{figure}[!hbtp]
  \centering
  \includegraphics[width=0.9\textwidth]{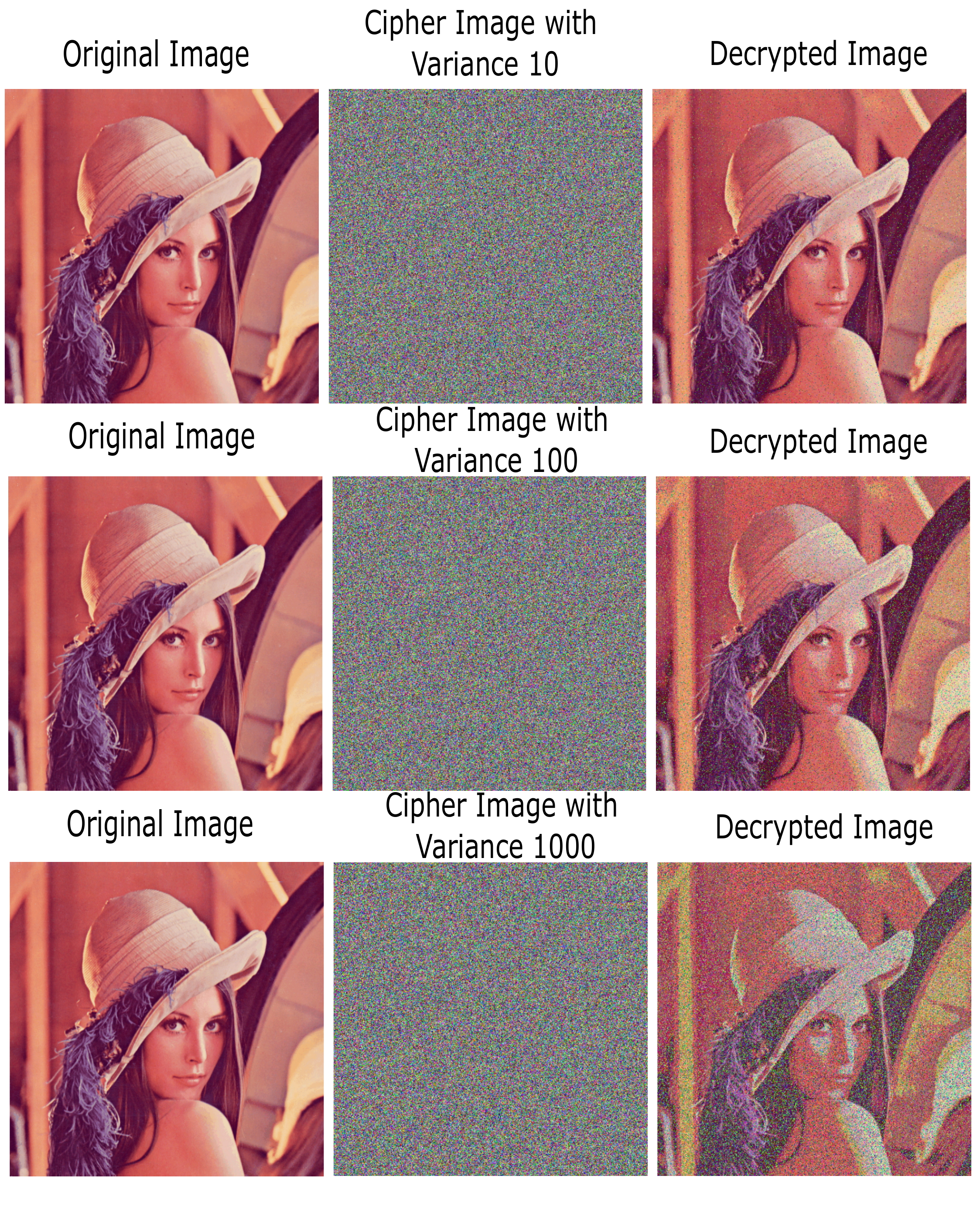}
  \caption{Encryption algorithm, which is not altered by the noise, is given in the figure. The first column is for the original image; the second one is for cipher images which are contaminated by noise at different intensities (10[first row], 100[second row], 1000[third row]); and the third column is used to represent the decrypted image after noise addition, which explains how well different levels of noise can be handled by this algorithm.}
  \label{rob to noise}
\end{figure}

During the noise robustness test, we study how noise-proof our encryption algorithm is at different levels of noise with which it can still operate even after such distortions that distort its input. In our method, we use a hyperchaotic map for the creation of sequences employed during the encryption or decryption of images. To be more specific, we test the algorithm's strength by injecting Gaussian noise of varying variances into the encrypted image before decrypting it again to check whether the original image was restored correctly. The parameters of the chaotic map, \(a_1\), \(a_2\), \(a_3\), \(b_1\), \(b_2\) and \(c\) define the hyperchaotic behaviour of this map and are thus necessary for producing its chaotic sequences. These sequences are, in turn, utilized in encrypting an image through transforming the initial pixel values by means of XOR operations as well as modulo arithmetic while an encrypted image gets exposed to Gaussian noise, imitating possible actual interferences happening in the world. In our analysis, there were variances of 10, 100 and 1000, respectively, representing low noise level, moderate noise level and low noise level, respectively (See figure\ref{rob to noise}). The added noise form was a Gaussian noise matrix by setting its variance at a certain level and superimposing it on the encrypted image pixels. The noisy encrypted image was decrypted with the same chaotic sequences, which was used to examine how robust the algorithm would be in such conditions. Despite having much higher noise levels, the decrypted images could be recognized, showing how strong this encryption was. Our noise analysis highlights the algorithm’s robustness regarding capacity under noise attacks that would otherwise compromise decrypted image integrity. This makes such encryption schemes reliable even under various noisy environments, thus positioning them as a robust option when it comes to sending secure images.
\begin{figure}[!hbtp]
  \centering
  \includegraphics[width=1.0\textwidth]{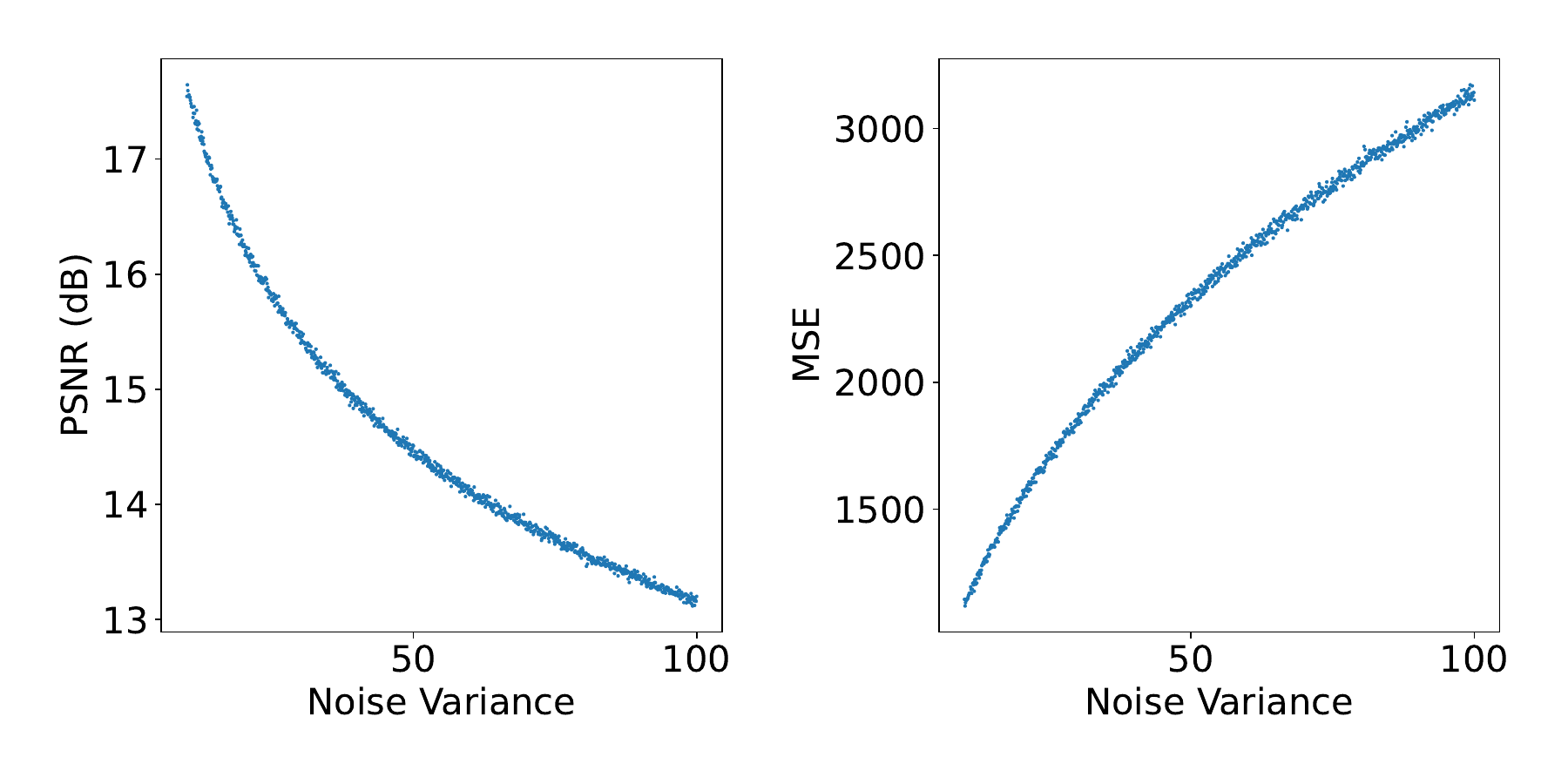}
  \caption{Robustness to noise analysis uncovers an inverse relationship between PSNR and MSE. The increase in noise variance leads to lower PSNR and higher MSE, indicating reduced image quality. This is a non-linear relationship implying that the effect of low levels of noise is quite minimal but increases significantly at higher levels.}
  \label{fig:7}
  
\end{figure}

\subsection{Differential Attack Analysis}
Differential attack analysis is a technique used to determine the strength of cryptographic algorithms, mainly in image encryption and decryption. It is a procedure done to analyze the variation between two forms of an image, an original image and a decrypted image, in order to disclose some weak points within the encryption algorithm. It tries to detect patterns or anomalies that may be utilized by an intruder to breach the security of the encryption scheme. The wide area of image encryption and decryption employs differential analysis for the attack of a particular algorithm. This helps the researchers to test the robustness of the encryption algorithms on the grounds of how much resistance they offer to differential attacks. These attacks simulate potential attacks from a possible attacker or enemy. From this point of view, the developed algorithms can be improved further by nullifying the weaknesses in the algorithms for the encryption of sensitive image data.

We subjected our encryption algorithm to a thorough analysis through differential attack. Original images were compared to their corresponding decrypted versions to test and determine how well the process of encryption retained image integrity and security. Using the results of the analysis, a table with quantitative information on the performance of our encryption algorithm was compiled (see Table\ref{anotable}. To conduct a differential attack analysis, we first encrypted the plain images with our encryption algorithm. Then, we decrypted the encrypted images to obtain the reconstructed images. We compared the pixel values of the original and decrypted images and analyzed the differences or deviations between them. It was quantified using metrics such as the Number of Pixel Change Rate (NPCR) and Unified Average Change Intensity (UACI), which quantify the similarity between the original and decrypted images. The table \ref{anotable} is a summary of the results we achieved by differential attack analysis in our research paper. It contains the NPCR and UACI values of each image analyzed, from which we can get an idea of the strength of our encryption algorithm in maintaining the image properties and security. The NPCR values represent the percentage of changed pixels between the original and the decrypted images, and the UACI values give the mean intensity of the changed pixels. The results of the differential attack analysis confirm the strength of our encryption and decryption process. The resultant NPCR is quite low and the UACI values are moderate, indicating that our encryption algorithm makes a significant improvement in image integrity and, at the same time, offers enough security against differential attacks. This is a sure indication of how reliable and efficient our encryption scheme is for the protection of sensitive image data from possible adversaries.

\begin{table}[htbp]
    \centering
    
    \label{tab:noise_analysis}
    \begin{tabular}{|c|c|c|}
        \hline
        \textbf{Image} & \textbf{NPCR (\%)} & \textbf{UACI} \\
        \hline
        Aeroplane & 99.90 & 68.78 \\
        Brain & 99.88 & 82.72 \\
        Lena & 99.76 & 47.66 \\
        Rainier & 99.86 & 62.34 \\
        \hline
    \end{tabular}
    \caption{Robustness to noise analysis for different images. From the NPCR (Number of Pixel Change Rate) and UACI (Unified Average Change Intensity) metrics, the results indicate the effectiveness of the noise-resistance property of these images. It is suggested that higher NPCR and lower UACI values lend better noise resistance.}
    \label{anotable}
\end{table}
\subsection{Key Sensitivity Analysis}
Key sensitivity analysis is a very important technique when talking about cryptosystems, especially in the case of image encryption and decryption. It deals with assessing the strength and security of encryption algorithms by scrutinizing the variations in the encryption key, which brings about variations in the secured information's security and integrity. It is sensitive to changes in cryptographic keys. In the area of image encryption and decryption, key sensitivity analysis is of prime importance to determine the robustness of encryption techniques against possible attacks or unauthorized access. Researchers can control the image encryption environment by systematically altering key parameters to understand the effect of key modification on the quality, security, and vulnerability of the encrypted image during decryption. In this paper, key sensitivity analysis was used to scrutinize the behaviour of a hyperchaotic encryption algorithm with slight changes in the parameters of the map under consideration. Specifically, the research looked into how changes in the key of 0.01 of some chaotic map coefficients affect the encryption and decryption of grayscale and coloured images. Analysis of the encrypted and decrypted images enabled knowing the sensitivity to key modification of the algorithm along with the robustness of the algorithm against potential attacks and decryptions. The original parameters of the hyperchaotic map were fixed so as to obtain extreme hyperchaotic signals. Then, a slight variation was brought about in these parameters to simulate a change in the encryption key. The image was first encrypted using the original key and then decrypted using the altered key parameters. The comparison of the resulting decryption with the original input image was performed to verify the key sensitivity of the decryption process. Key sensitivity analysis is generally an illustration of the key sensitivity of the algorithm, providing important insights into the corresponding security implications. At the same time, it has been shown that the rise of robust algorithms for data security is of prime significance in view of key modifications that are important for real-time applications. The work also illustrated the importance of thorough testing and evaluation methodologies for the key sensitivity analysis, determining the efficiency and reliability of the cryptographic system in real-world applications.
\begin{figure}[!hbtp]
  \centering
  \includegraphics[width=0.9\textwidth]{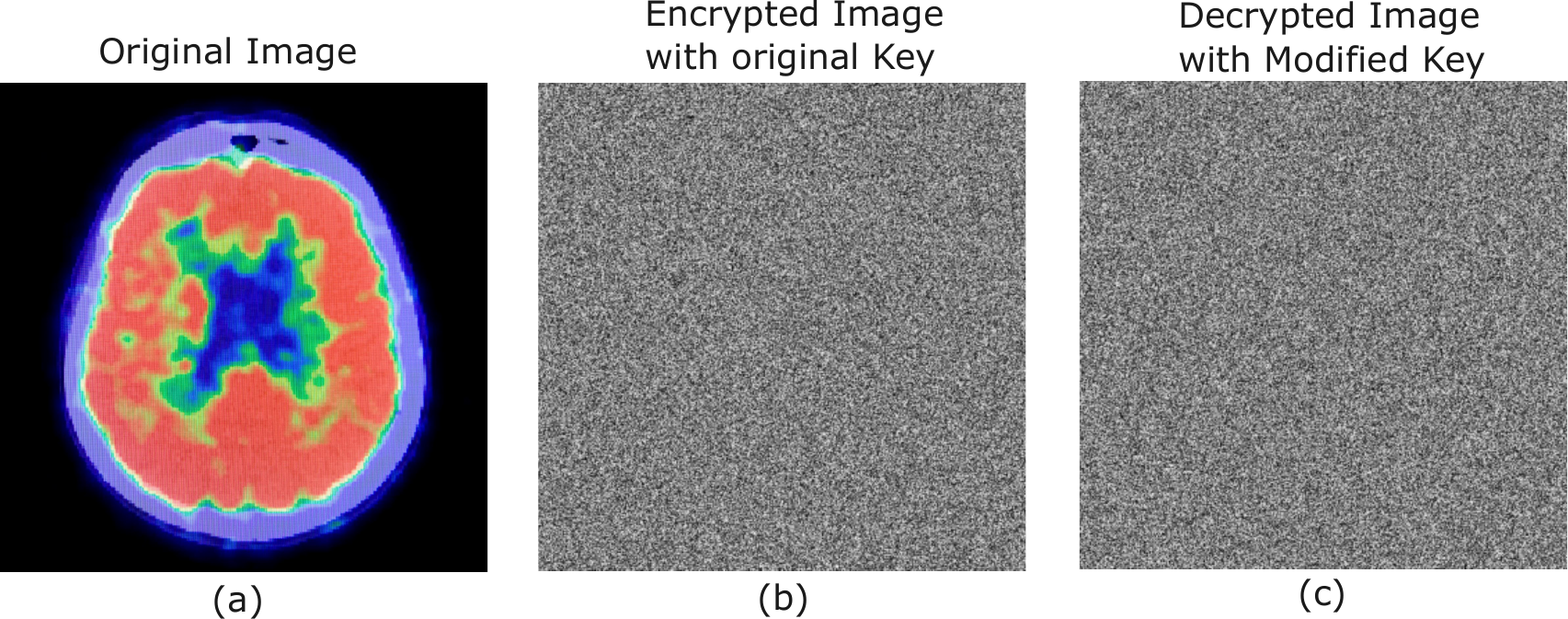}
  \caption{This image shows the working of a symmetric-key encryption algorithm in securing the medical image of the brain. In (a) shows the original medical image of the brain. In (b), shows the image that was encrypted with the original key. It is translated into a noisy random pattern so that it is not easily recognized. However, when encryption is attempted with a key that is slightly modified, i.e. (c), the decryption fails, and an image is produced with random distortion that is not recognizable. This emphasizes the need for secure storage of the encryption key because even slight modification will render the data unreadable.}
  \label{fig:}
\end{figure}

\section{Results} \label{section7}

In this paper, we take the value of the parameters \(a_1 = 0.03\), \(a_2 = 0.25\), \(a_3 = 0.11\), \(b_1 = 4\), \(b_2 = 1.2\), and \(c = 2.15\). We see that the system presents characteristics of high sensitivity to initial conditions and the existence of many attractors which is identified as dynamic instability. When \(a_1\) was adjusted to 0.05, the behaviour of the system changed significantly while the parameters were constant; this validates the fact that the system is both delicate and sensitive. This should lead to an effective use of the properties of the hyperchaotic map in our encryption algorithm.
    
We analyzed the behavior of the system for values of \(k = 1.74\), \(k = 1.75\), and \(k = 1.76\). Each plot showed that when a system goes through a transition from periodic to chaotic regimes, it displays a more chaotic behaviour for \(k = 1.75\), but in each case, the most chaotic display was taken as the one that presented the correct graph. In this way, our encryption algorithm would experience the complexity of the memristor map and would ensure it is more secure and resistant to predictable patterns.
    
We effectively encrypted images with highly scrambled pixel values using the 3D hyperchaotic map and 2D memristor map combined with a Convolutional Neural Network (CNN), producing very complex encrypted images. During decryption, it reconstructed the original images properly because the same chaotic sequences and CNN were used. Hence, our proposed encryption scheme is made efficient in both the encryption and decryption processes, and it preserves the integrity of the images.
    
We attempted to make the generation of cryptographic keys as random and complex as possible by using a cryptographically secure random number generator. This ensures that our encryption algorithm will be secured for privacy and safe against unauthorized access.
The result of the entropy of the encrypted images was approximately 7.598. This value is sufficiently high, meaning that the randomness and complexity are well maintained, hence providing the security features this algorithm has to offer to the hidden information in the original images.
The correlation analysis indicates a successful decryption occurrence between plaintext and ciphertext channels through positive coefficients, while it indicates a successful encryption occurrence between plaintext and original channels through negative coefficients. The obtained correlation coefficients prove that good decryption and encryption are happening as the values approach 1.0, and a successfully encrypted image is one which is hard to extract as the values approach 0, hence confirming the applicability of the proposed method.
For anomaly detection, we got a low MSE and high SSIM, showing that the encryption and decryption process keep the quality and integrity of the image. This makes the algorithm effective in delivering the important features of the original image as well.
Results of decryption on images with added Gaussian noise prove that the encryption method is robust and suitable in real noisy conditions. This indicates that our encryption mechanism is robust even in adverse conditions.
The result of differential attack analysis presented a low Pixel Change Rate (NPCR) and a fair Unified Average Change Intensity (UACI) value, indicating a good resistance of the algorithm against the differential attack. That proves our encryption algorithm to be steady and reliable when facing these cryptographic issues.

\section{Conclusion} \label{section8}
In this paper, an image encryption and decryption method with a 3D hyperchaotic map and a 2D memristor map is proposed, whose security is strengthened by a deep neural network. Our design generates intricate encrypted sequences that are dynamically unstable and highly sensitive to initial conditions, thus facilitating rigorous parameter analysis and system behaviour testing. This algorithm ensures high-quality decryption. It maintains robust encryption because of the hyperchaotic properties of the 3D map, which secure the data, while during decryption, the CNN maintains the integrity of the image. Our key findings include that these methods realize large entropy and low correlation coefficients of the encrypted images, indicating strong resistance to cryptographic attacks. Moreover, the system performs reliably due to its resilience against Gaussian noise and differential attacks, proving it to be robust and reliable. The practical implications are very important in developing an image-encryption method that is highly secure and efficient and can protect sensitive visual data communicating on a digital basis. However, a lot of optimization-related problems with regard to the computational efficiency and scalability of the algorithm are yet to be sorted out. Future research will further extend this method to colour images and video encryption using different chaotic and hyperchaotic systems and advanced neural network architectures. Security strengthening can be done with quantum computing techniques. This has to be implemented on a real-time basis and should be compared to other encryption algorithms available. Continuous refinement will validate our approach and work towards betterment. Our method will keep pace with the changing technologies in image encryption to offer robust security solutions, protecting organizational reputation in our increasingly digital world.

\section*{Conflict of interest}
 The authors declare that they have no conflict of interest.

 \section*{Data Availability Statement}
 The data that support the findings of this study are available within the article.

\bibliographystyle{unsrt}
\bibliography{Arxiv}

\end{document}